\documentclass{cpbtex}
\usepackage{CJK}

\usepackage{CJK}
\expandafter\let\csname equation*\endcsname\relax
\expandafter\let\csname endequation*\endcsname\relax
\usepackage{amsmath}
\usepackage{amsthm,amsmath,amssymb}
\usepackage{mathrsfs}
\usepackage{amssymb}
\usepackage{hyperref}
\usepackage{cite}
\usepackage{indentfirst}
\usepackage{graphicx}
\usepackage{subfigure}
\usepackage{epstopdf}
\usepackage{subfigure}
\usepackage[font=small,labelfont=bf,labelsep=none]{caption}
\usepackage{amsmath}
\usepackage{hyperref}
\usepackage{cite}
\usepackage{indentfirst}
\usepackage{graphicx}
\usepackage{epstopdf}
\usepackage{subfigure}
\usepackage{array}
\usepackage[font=small,labelfont=bf,labelsep=none]{caption}
\begin{document}
	
	\title{Conservation law and Lie symmetry analysis of the (1+1) dimensional dispersive long-wave equation}
	
	% Title should be concise; avoid abbreviations if possible; and not begin with `A', `An', `The', or `Study on'.
	
	\author{Long Ju$^{1}$, Faiza Afzal$^{1}$ and Yufeng Zhang$^{1,2,}$\thanks{Corresponding author. E-mail:zhangyfcumt@163.com}\\
		$^{1}${School of Mathematics,China University of Mining and Technology, Xuzhou, 221116, China;}\\
$^{2}${Jiangsu Center for Applied Mathematics (CUMT), Xuzhou, Jiangsu 221116, China.} \\
	  % The line break was forced via \\% The line break was forced via \\
	}   % The line break was forced via \\
	% 1. For Chinese authors, the name in Chinese characters should also be given. For example, Gang Liu(Áõ¸Õ), Xiao-Ming Li(ÀîÏþÃ÷)
	% 2. Please ensure that every author approves the submission of the manuscript
	% 3. Abbreviations should not be used in the affiliations
	
	\date{\today}
	\maketitle
	\begin{abstract}
		In this paper, we mainly study the integrability of (1+1)-dimensional dispersive long-wave equation. Firstly, the Lie symmetry analysis of the equation is carried out in the first part. And the optimal system of the equation is obtained according to the symmetry, and the invariant solution and the reduced form of the target equation are solved according to the results. Secondly, we use different methods to solve the conservation law of the target equation. To begin with, we give the adjoint determination equation and adjoint symmetry of the (1+1)-dimensional dispersive long-wave equation, and use the adjoint symmetry as the equation multiplier to find several conservation laws. Then we get a Lie bracket by using the relationship between the symmetry of the equation and the adjoint symmetry. Next its strict self-adjoint property is verified, and its conservation laws are solved by Ibragimov's method. Finally, the conservation laws of the target equation are solved by Noether's theorem. Thirdly we calculate some exact solutions of the target equation by three different methods. In the end of the paper, the Hamiltonian structure of the target equation, the generalized pre-symplectic that maps symmetries into adjoint-symmetries and some of its soliton solutions are calculated. In conclusion, we use the direct construction of conservation law method, Ibragimov's method and so on to solve some new conservation laws of (1+1)-dimensional dispersive long-wave equation, use the relationship between symmetry and adjoint symmetry to construct the corresponding Lie brackets, and obtain some linear soliton solutions according to the conservation law of the equation.
		~
	\end{abstract}

	\section{Introduction}
	Lie symmetry has been applied in most fields of mathematics and has played a crucial role, such as differential equations~\cite{ref1,ref2}, Lie algebra~\cite{ref3,ref4} and classical mechanics~\cite{ref5}. Meanwhile, the Lie symmetry method is also considered to be one of the most effective methods to derive the explicit solutions of nonlinear partial differential equations. At the same time, on the basis of symmetry, the integrability of differential equations can be considered. From an algebraic viewpoint, the infinitesimal symmetries of a PDE are the solutions
	of the linearisation (Frechet derivative) equation holding on the space of solutions to the partial differential equation. Solutions of the adjoint linearisation equation, holding on the space of solutions to the partial differential equation, are called adjoint-symmetries and provide a direct link to conservation laws. In particular, adjoint-symmetries
	that satisfy a certain variational condition represent multipliers which yield conservation laws.
	\par In the past study of differential equations, the conservation law is undoubtedly a very important
	part, especially in terms of integrability and linearization~\cite{ref6,ref7,ref8}. How to solve the conservation law of the
	equation has become a first problem to be faced. Conservation laws are indispensable in the study of mathematical physics equations. At present, there are many methods to construct the conservation laws of mathematical physics equations, such as 
	Noether theorem, known conservation laws of symmetric action, the variational derivative method, the recursive formula method and so on. In this study, we choose three methods to solve the conservation laws of partial differential equations, which are direct construction of conservation laws~\cite{ref9,ref10,ref11,ref12}, adjoint equation method~\cite{ref13,ref14,ref15,ref16,ref17} and using Noether's theorem to solve conservation laws.
	\par This article manly deals with the (1+1) dimensional dispersive long-wave equation~\cite{ref19} of the form 
	\begin{equation}
		\begin{aligned}
			&u_t+uu_x+v_x=0,\\
			&v_t+u_xv+uv_x+\frac{1}{3}u_{xxx}=0,
		\end{aligned}
	\end{equation}
where $v$ is the elevation of water and $u$ is the surface velocity of water along $x$-direction. System (1) can
be traced back to the works of Broer~\cite{ref20}, Kaup~\cite{ref21},
Martinez~\cite{ref22} and Kupershmidt~\cite{ref23}. It is very useful
for coastal and civil engineers to comprehend the solutions of the nonlinear water wave model in harbor and coastal designs.
In this paper, we mainly use three ways to solve the conservation law of equation (1) and analyze its Lie symmetry. At the same time, we use the relationship between symmetry and adjoint symmetry~\cite{ref24,ref25,ref26,ref27,ref28} to obtain a new bracket of the target equation, and solve its Hamiltonian structure and linear soliton solution. Finally, we calculate some exact solutions~\cite{ref29,ref30,ref31,ref32,ref33} of the target equation by the extended Kudryashov method and seek a few solutions in terms of hyperbolic tangent functions and $\exp(\xi)$.
	
\section{Lie symmetries analysis}
\subsection{Lie point symmetries}
In this section, Lie symmetry analysis on the (1+1) dimension dispersive long-wave equation was performed alongside considering one Lie 
group of transformations parameter
\begin{equation}
	\begin{aligned}
		&t\rightarrow t+\epsilon\xi^{1}(x,t,u,v),
		x\rightarrow x+\epsilon\xi^{2}(x,t,u,v),\\
		&u\rightarrow u+\epsilon\xi^{3}(x,t,u,v),
		v\rightarrow v+\epsilon\eta^{1}(x,t,u,v),
	\end{aligned}
\end{equation}
with a small parameter $\epsilon \leq 1$. And the corresponding generator of the Lie algebra of the form
\begin{equation}
	X=\xi^{1}(t,x,u,v)\frac{\partial}{\partial t}+\xi^{2}(t,x,u,v)\frac{\partial}{\partial x}+\eta^{1}(t,x,u,v)\frac{\partial}{\partial u}+\eta^{2}(t,x,u,v)\frac{\partial}{\partial v}.
\end{equation}
Thus the third prolongation $pr^{(3)}X$ is
\begin{equation}
	pr^{(3)}X=X+\eta^{1}_{t}\frac{\partial}{\partial u_{t}}+\eta^{1}_{x}\frac{\partial}{\partial u_{x}}+\eta^{2}_{t}\frac{\partial}{\partial v_{t}}+\eta^{2}_{x}\frac{\partial}{\partial v_{x}}+\eta^{1}_{xx}\frac{\partial}{\partial u_{xx}}+\eta^{1}_{xy}\frac{\partial}{\partial u_{xy}}+\cdots.
\end{equation}
where $\eta^{\alpha}_{i}=D_{i}\eta^{\alpha}-(D_{i}\xi^{j})u_{j},i=1,2,\alpha=1,2.$ $\eta^{\alpha}_{i_{i}i_{2}\cdots i_{k}}=D_{i_{k}}\eta^{\alpha}_{i_{1}i_{2}\cdots i_{k-1}}-(D_{i_{k}}\xi^{j})u_{i_{1}i_{2}\cdots i_{k-1}j},i_{l}=1,2,\alpha=1,2$ for $l=1,2,\dots k$ with $k=2,3,\dots.$
and the operators $D_{x}$ and $D_{t}$ are the total derivatives with respect to $x$ and $t$. The determining equations of Eq. (1) arises
from the following invariance condition
\begin{equation}
	pr^{(3)}X(G)|_{G=0}=0.
\end{equation}
		Then we obtain the overdetermined system of	the partial differential equations
	\begin{equation}
		\begin{aligned}
			&\xi^1_{tt}=0,\xi^2_{tt}=0,\xi^1_u=0,\xi^1_v=0,\xi^1_x=0,\xi^2_u=0,\\
			&\xi^2_v=0,\xi^2_x=\frac{\xi^1_t}{2},\eta^1=\frac{u\xi^1_t}{2}+\xi^2_t,\eta^2=v\xi^1_t.
		\end{aligned}
	\end{equation}
Solving the system, one can get
\begin{equation}
	\begin{aligned}
		\xi^1=C_1t+C_2,\xi^2=\frac{C_1}{2}x+C_3t+C_4,\eta^1=\frac{C_1u}{2}+C_3,\eta^2=C_1v,			
	\end{aligned}
\end{equation}
where $C_i,i=1,\cdots,4.$ are arbitrary constants. Hence the infinitesimal symmetries of Eq. (1) form the four dimensional
Lie algebra $L_6$ spanned by the following independent operators:
\begin{equation}
	\begin{aligned}
		&X_{1}=\frac{\partial}{\partial t},X_{2}=\frac{\partial}{\partial x},X_{3}=t\frac{\partial}{\partial x}+\frac{\partial}{\partial u},\\
		&X_{4}=\frac{x}{2}\frac{\partial}{\partial x}+t\frac{\partial}{\partial t}-\frac{u}{2}\frac{\partial}{\partial u}-v\frac{\partial}{\partial v}.\\
	\end{aligned}
\end{equation}
By solving the following system of ordinary differential equations with initial condition
\begin{equation}
	\begin{aligned}
		&\frac{dt^{*}}{d \epsilon}=\xi^{1}(t^{*},x^{*},u^{*},v^{*})|_{\epsilon=0}=t,	\frac{dx^{*}}{d \epsilon}=\xi^{2}(t^{*},x^{*},u^{*},v^{*})|_{\epsilon=0}=x,\\
		&\frac{du^{*}}{d \epsilon}=\eta^{1}(t^{*},x^{*},u^{*},v^{*})|_{\epsilon=0}=u,	\frac{dv^{*}}{d \epsilon}=\eta^{2}(t^{*},x^{*},u^{*},v^{*})|_{\epsilon=0}=v.
	\end{aligned}
\end{equation}
We can obtain the group transformation which is generated by the infinitesimal generator $X_{i}(i=1,\dots,5)$, respectively
\begin{equation}
	\begin{aligned}
		&G_{1}:(t,x,u,v)\rightarrow(t+\epsilon,x,u,v),\\
		&G_{2}:(t,x,u,v)\rightarrow(t,x+\epsilon,u,v),\\
		&G_{3}:(t,x,u,v)\rightarrow(t,x+t\epsilon,u+\epsilon,v),\\
		&G_{4}:(t,x,u,v)\rightarrow(te^{\epsilon},xe^{\frac{\epsilon}{2}},u=ue^{-\frac{\epsilon}{2}},v=ve^{-\epsilon}).
	\end{aligned}
\end{equation}
~\\
\textbf{Theorem 1.} If {$(u(x,t),v(x,t))$} is a solution of Heisenberg equation, so are the functions
\begin{equation}
	\begin{aligned}
		&G_1(\epsilon) \cdot u(x,t)=u(x,t-\epsilon),G_1(\epsilon) \cdot v(x,t)=v(x,t-\epsilon),\\
		&G_2(\epsilon) \cdot u(x,t)=u(x-\epsilon,t),G_1(\epsilon) \cdot v(x,t)=v(x-\epsilon,t),\\
		&G_3(\epsilon) \cdot u(x,t)=u(x-t\epsilon,t)-\epsilon,G_3(\epsilon) \cdot v(x,t)=v(x-t\epsilon,t),\\
		&G_4(\epsilon) \cdot u(x,t)=u(xe^{-\frac{\epsilon}{2}},te^{-\epsilon})e^{\frac{\epsilon}{2}},
		G_4(\epsilon) \cdot v(x,t)=v(xe^{-\frac{\epsilon}{2}},te^{-\epsilon})e^{\epsilon}.
	\end{aligned}
\end{equation}

\subsection{Optimal system of one-dimensional subalgebras}
In 2010, Ibragimov et al.
presented a concise method to get the optimal system, which only relies on the commutator table.
Then Abdulwahhab constructed an optimal system of the 2-dimensional Burgers equations.
In this part, we try to find the optimal system of (1+1)-dimensional dispersive long-wave equation. First, we construct a Lie bracket $L_5$.
Their algebra has the non-zero commutators
\begin{equation}
	[X_1,X_3]=X_2, [X_1,X_4]=X_1, [X_2,X_4]=\frac{1}{2}X_2, [X_3,X_4]=-\frac{1}{2}X_3.
\end{equation}
We take any operator $X=l^1X_1+l^2X_2+l^3X_3+l^4X_4.$
To find the linear transformations of the vector $l=(l^1,l^2,l^3,l^4)$, we can use the following generators
\begin{equation}
	E_{i}=c_{ij}^{k}l^j\frac{\partial}{\partial l^k},i=1,2,3,4,
\end{equation}
where $c_{ij}^k$ is defined by the formula $[X_i.X_j]=c_{ij}^kX^k.$ So after calculation, $E_1,E_2,E_3,E_4$  has the following forms:
\begin{equation}
	\begin{aligned}
		&E_1=l^3\frac{\partial}{\partial l_2}+l^4\frac{\partial}{\partial l^1},E_2=\frac{1}{2}l^4\frac{\partial}{\partial l^2},\\
		&E_3=-l^1\frac{\partial}{\partial l^2},E_4=-l^1\frac{\partial}{\partial l^1}-\frac{1}{2}l^2\frac{\partial}{\partial l^2}-\frac{1}{2}l^3\frac{\partial}{\partial l^3}.
	\end{aligned}
\end{equation}
Let us find the transformations provided by the generators (14). 
For the generator $E_1,E_2,E_3,E_4$, the Lie equations with the parameter $a_1,a_2,a_3,a_4$ are written
\begin{equation}
	\begin{aligned}
		&\frac{d\widetilde{l}^1}{a_1}=\widetilde{l}^4,\frac{d\widetilde{l}^2}{a_1}=\widetilde{l}^3,\frac{d\widetilde{l}^3}{a_1}=0,\frac{d\widetilde{l}^4}{a_1}=0,\\
		&\frac{d\widetilde{l}^1}{a_2}=0,\frac{d\widetilde{l}^2}{a_2}=\frac{1}{2}\widetilde{l}^4,\frac{d\widetilde{l}^3}{a_2}=0,\frac{d\widetilde{l}^4}{a_2}=0,\\
		&\frac{d\widetilde{l}^1}{a_3}=0,\frac{d\widetilde{l}^2}{a_3}=-\widetilde{l}^1,\frac{d\widetilde{l}^3}{a_3}=-\frac{1}{2}\widetilde{l}^4,\frac{d\widetilde{l}^4}{a_3}=0,\\
		&\frac{d\widetilde{l}^1}{a_4}=-\widetilde{l}^1,\frac{d\widetilde{l}^2}{a_4}=-\frac{1}{2}\widetilde{l}^2,\frac{d\widetilde{l}^3}{a_4}=-\frac{1}{2}\widetilde{l}^3,\frac{d\widetilde{l}^4}{a_4}=0.
	\end{aligned}
\end{equation}
The solutions of these equations give the transformations
\begin{equation}
	\begin{aligned}
		&T_1:\widetilde{l}^1=l_1+a_1l^4,\widetilde{l}^2=l^2+a_1l^3,\widetilde{l}^3=l^3,\widetilde{l}^4=l^4,\\
		&T_2:\widetilde{l}^1=l^1,\widetilde{l}^2=l^2,\widetilde{l}^3=l^3+\frac{1}{2}a_2l^4,\widetilde{l}^4=l^4,\\
		&T_3:\widetilde{l}^1=l^1,\widetilde{l}^2=l^2-a_3l^1,\widetilde{l}^3=l^3-\frac{1}{2}a_3l^4,\widetilde{l}^4=l^4,\\
		&T_4:\widetilde{l}^1=e^{-a_4}l^1,\widetilde{l}^2=e^{-\frac{1}{2}a_4}l^2,\widetilde{l}^3=e^{-\frac{1}{2}a_4}l^3,\widetilde{l}^4=l^4.
	\end{aligned}
\end{equation}
The construction of the optimal system requires a simplification of the vector $l=(l^1,l^2,l^3,l^4)$ by use of $T_1,T_2,T_3,T_4$. And the construction will be divided into two cases.
~\\
\textbf{Case 1.} $l^1\neq0.$
By taking $a_3=\frac{l^2}{l^1}$ in $T_3$, we can make $\widetilde{l}^2=0$. The vector is reduced to the form
$l=(l^1,0,l^3,l^4)$.~\\
\textbf{1.1} $l^4\neq0$
By taking $a_2=-2\frac{l^3}{l^4}$ in $T_2$, we can make $\widetilde{l}^3=0$. The vector is reduced to the form
$l=(l^1,0,0,l^4)$.~\\
Then by taking $a_1=-\frac{l^1}{l^4}$.The vector is reduced to the form
$l=(0,0,0,l^4)$.~\\ we obtain the following representatives
$$X_4.$$
\textbf{1.2} $l^4\neq 0$ 
The vector is reduced to the form
$l=(l^1,0,l^3,0)$.~\\
Taking all the possible combinations, we obtain the following representatives
$$X_1,X_1\pm X_3.$$
\textbf{Case 2.} $l^1=0.$
 The vector is reduced to the form
$l=(0,l^2,l^3,l^4)$.~\\
\textbf{2.1} $l^3\neq 0$. By taking $a_1=-\frac{l^2}{l^3}$ in $T_1$, we can make $\widetilde{l}^2=0$. The vector is reduced to the form
$l=(0,0,l^3,l^4)$.~\\
Then if we set $l^4=0$, we obtain the following representatives
$$X_3.$$
If $l^4\neq0$, by taking $a_3=\frac{2l^3}{l^4}$ in $T_3$. The vector is reduced to the form
$l=(0,0,0,l^4)$. we obtain the following representatives
$$X_4.$$
\textbf{2.2} $l^3=0$
The vector is reduced to the form
$l=(0,l^2,0,l^4)$. Taking all the possible combinations, we obtain the following representatives
$$X_2,X_4,X_2\pm X_3.$$
\textbf{Theorem 2.} The following operators provide an optimal system of one-dimensional subalgebras of the Lie algebra spanned by Eq. (1):
$$X_1,X_2,X_3,X_4,X_1\pm X_3,X_2\pm X_4$$.
\subsection{Reductions and the invariant solutions}
After we calculate the optimal system of subalgebras, we can investigate the symmetry reductions of Eq. (1) by 
integrating the characteristic equations. Next, we analyze the reduced order and invariant solution of the Eq. (1).
\subsubsection{Solutions through $X_{1}+X_3$}
For the generator $$X_{1}+X_3=\frac{\partial}{\partial t}+t\frac{\partial}{\partial x}+\frac{\partial}{\partial u},$$ the characteristic equation is written as
\begin{equation}
	\frac{dt}{1}=\frac{dx}{t}=\frac{du}{1},
\end{equation}
We can get a similar variables according to the characteristic equation:
$$Q=\frac{t^2}{2}-x,u=t+f(Q),v=g(Q).$$
After bringing them into Eq.(1), we can obtain the reduced form of it:
\begin{equation}
	\begin{aligned}
		&1+f^{'}t+(t+f)(-f^{'})-g^{'}=0,\\
		&tg^{'}-f^{'}g+(t+f)(-g^{'})-\frac{f^{'''}}{3}=0.
	\end{aligned}
\end{equation}
By solving it, We can get the form of the solution of the Eq. (1)
$$f=c_1,g=Q+c_2=\frac{t^2}{2}-x+c_2.$$
So we obtain the invaint solution of Eq. (1)
\begin{equation}
	u=t+c_1,v-\frac{t^2}{2}-x+c_2,
\end{equation}
where $c_1$ and $c_2$ are arbitrary constants.
\subsubsection{Solutions through $X_{2}+X_4$}
For the generator $$X_{2}+X_4=t\frac{\partial}{\partial t}+(\frac{x}{2}+1)\frac{\partial}{\partial x}-\frac{u}{2}\frac{\partial}{\partial u}-v\frac{\partial}{\partial v},$$ the characteristic equation is written as
\begin{equation}
	\frac{dx}{1+\frac{x}{2}}=\frac{dt}{t}=\frac{du}{-\frac{u}{2}}=\frac{dv}{-v}.
\end{equation}
By calculation, we can get the expressions of similar variables and $u,v$:
$$R=\frac{x+2}{\sqrt{t}},u=\frac{f(R)}{\sqrt{t}},v=\frac{g(R)}{t}.$$
Then we can get the reduced form of Eq. (1)
\begin{equation}
	\begin{aligned}
	&	-R\frac{f^{'}}{2}-\frac{f}{2}+ff^{'}+g^{'}=0,\\
		&-g-R\frac{g^{'}}{2}+gf^{'}+fg^{'}+\frac{f^{'''}}{3}=0.
	\end{aligned}
\end{equation}
By solving the system (21), we can obtain the solution of it
$$f=R,g=c_1.$$
So we get the invarint solution of Eq. (1)
\begin{equation}
	u=\frac{1}{\sqrt{t}}\cdot \frac{x+2}{\sqrt{t}}=\frac{x+2}{t},v=\frac{c_1}{t},
\end{equation}
where $c_{1}$ is an arbitrary constant. 
\section{Adjoint symmetries analysis}
\subsection{Constructing conservation laws by adjoint symmetries}
In this section, we try to use the adjoint determination equation of the equation to solve its adjoint symmetry, and use the adjoint symmetry to solve the conservation law of the target equation. First we consider a system of partial differential equations with $N$ independent variables $\boldsymbol{u}=(u^{1},\dots,u^{N})$ and $n+1$ independent variables $(t,\boldsymbol{x})$, the forms are as follows:
\begin{equation}
	G^{i}=\frac{\partial u^{j}}{\partial t}+g^{j}(t,\boldsymbol{x},\boldsymbol{u},\partial_{\boldsymbol{x}}\boldsymbol{u},\dots,\partial^{m}_{\boldsymbol{x}}\boldsymbol{u})=0,j=1,\dots,N,
\end{equation}
with $\boldsymbol{x}$ derivatives of $\boldsymbol{u}$ up to some order $m$. And we can use $\partial_{\boldsymbol{x}}\boldsymbol{u},\partial_{\boldsymbol{x}}^{2}\boldsymbol{u}$, etc. to represent all the derivatives of $u^{j}$ with respect to ${x^{i}}$. We denote partial derivatives $\frac{\partial}{\partial t}$ and $\frac{\partial}{\partial x^{i}}$
with $\boldsymbol{x}$ derivatives of $\boldsymbol{u}$ up to some order $m$. We denote partial derivatives $\frac{\partial}{\partial t}$ and $\frac{\partial}{\partial x^{i}}$
by subscripts $t$ and $i$ respectively. And likewise, $D_{t}$ and $D_{i}$ represent the total derivatives with respect to $x^{i}$ and $t$. 
~\\
\textbf{Definition 4} Multipliers for PDE system (23) are a set of expressions
$${\lambda_{1}(t,\boldsymbol{x},\boldsymbol{u},\partial_{\boldsymbol{x}}\boldsymbol{u},\dots,\partial^{q}_{\boldsymbol{x}}\boldsymbol{u}),\cdots,\lambda_{N}(t,\boldsymbol{x},\boldsymbol{u},\partial_{\boldsymbol{x}}\boldsymbol{u},\dots,\partial^{q}_{\boldsymbol{x}}\boldsymbol{u})}$$
satisfying
\begin{equation}
	(u^{j}_{t}+g^{j})\lambda_{j}=D_{t}\phi^{t}+D_{i}\phi^{i},
\end{equation}
for some expression $\phi^{t}(t,\boldsymbol{x},\boldsymbol{u},\partial_{\boldsymbol{x}}\boldsymbol{u},\dots,\partial^{k}_{\boldsymbol{x}}\boldsymbol{u})$ and 
$\phi^{i}(t,\boldsymbol{x},\boldsymbol{u},\partial_{\boldsymbol{x}}\boldsymbol{u},\dots,\partial^{k}_{\boldsymbol{x}}\boldsymbol{u})$ for all
functions $\boldsymbol{u}(t,\boldsymbol{x}).$
~\\
\textbf{Theorem 3}
For the Cauchy-Kovalevskaya PDE system (23), every nontrivial conservation
law in normal form $D_t\phi^t(\boldsymbol{x},t,\partial_{\boldsymbol{x}}\boldsymbol{u},\cdots)+D_i\phi^i(\boldsymbol{x},t,\partial_{\boldsymbol{x}}\boldsymbol{u},\cdots)=0$ is uniquely characterized by a set of multipliers $\Lambda_\sigma$ with no
dependence on ut and differential consequences, satisfying the relations $D_t\phi^t+D_i(\phi^i-\Gamma^i)=(u_t^\sigma+g^\sigma)\Lambda_\sigma$ and $\Lambda_{\sigma}=\widehat{E}_u^\sigma(\phi^t),\sigma=1,\cdots,N.$
holding for all functions $\boldsymbol{u}(t,\boldsymbol{x})$, where $\Gamma^i$ is given by an expression proportional to $u_t^\sigma+g^\sigma$ (and differential consequences), $\widehat{E}_u^\sigma=\partial_{u^\sigma}-D_i\partial u^\sigma_i+D_iD_j\partial_{u_{ij}^\sigma}+\cdots.$
\par
The standard determining condition for multiplier expressions $\Lambda_\sigma(t,\boldsymbol{x},\boldsymbol{u},\partial_{\boldsymbol{x}}\boldsymbol{u},\cdots)$ arises from the definition (24) by the wellknown result that divergence expressions are characterized by annihilation under the full
Euler operator
$$E_{u^\sigma}=\partial_{u^\sigma}-D_i\partial_{u_i^\sigma}-D_t\partial_{u_t^{\sigma}}+D_iD_j\partial_{u_{ij}^{\sigma}}+D_tD_j\partial_{u_{ij}^\sigma}+\cdots.$$
This yields
\begin{equation}
	E_{u^{\sigma}}(u_t^\rho\Lambda_{\rho}+g^\rho\Lambda_{\rho})=0.
\end{equation}
To carry out the splitting of $D_t\Lambda_{\sigma}$, we ues the identity
$$D_t=\partial_t+(u_t^\rho+g^\rho)\partial_{u^\rho}+(u_{ti}^\rho+D_ig^\rho)\partial_{u^\rho_i}+\cdots.$$
Consequently, the non-leading terms in Eq. (25) are as follows
\begin{equation}
	\begin{aligned}
	&-\frac{\partial\Lambda_{\sigma}}{\partial t}+(\frac{\partial\Lambda_{\sigma}}{\partial u^\rho}g^\rho+\frac{\partial\Lambda_{\sigma}}{\partial u^\rho_i}D_ig^\rho+\cdots+\frac{\partial\Lambda_{\sigma}}{\partial u^\rho_{i_1\cdots i_p}}D_i\cdots D_{i_p}g^\rho)\\
	&+\frac{g^\rho}{\partial u^\sigma}\Lambda_{\rho}-D_i(\frac{g^\rho}{\partial u^\sigma_i}\Lambda_{\rho})+\cdots+(-1)^mD_{i_1}\cdots D_{i_m}(\frac{g^\rho}{\partial u^\sigma_{i_1\cdots i_m,}}\Lambda_{\rho})=0,\sigma=1,\dots,N.
	\end{aligned}
\end{equation}
It is obviously the adjoint determining equation of Eq. (23), when $\boldsymbol{u}(t,\boldsymbol{x})$ is the solution of Eq.(23). Therefore, we know that the multiplier of the conservation law of the target equation is actually its special adjoint symmetry.
Next, we try to use the adjoint symmetry of Eq. (1) to solve its conservation law. First, we can write the adjoint determining equation of the Eq. (1)
\begin{equation}
	\begin{aligned}
			&-D_tQ_1-uD_xQ_1-vD_xQ_2-\frac{1}{3}D_x^3Q_2=0,\\
		&-D_tQ_2-uD_xQ_2-D_xQ_1=0,
	\end{aligned}
\end{equation}
where $Q_i,i=1,2.$ It has the expression $Q_i(x,t,\boldsymbol{u},\partial_{x}\boldsymbol{u},\partial_{x^2}\boldsymbol{u})$. By expanding the formula (27) and separating the coefficients, we can get an overdetermined equation system about $Q_i$. After solving the overdetermined equation system, we can obtain the adjoint symmetry of Eq. (1)
\begin{equation}
	\begin{aligned}
	&Q_1=(v_{xx}+\frac{9}{4}v^2+\frac{9}{4}u^2v+\frac{6uu_{xx}+3u_x^2}{4},u_{xx}+\frac{3u^3}{4}+\frac{9}{2}uv),\\
	&Q_2=(tv,tu-x),Q_3=(tv,\frac{u^2}{2}+v),Q_4=(v,u),Q_5=(1,0),Q_6=(0,1).
	\end{aligned}
\end{equation}
By verifying the folumar (25), we can find that $Q_i,i=1,\dots$ is the multiplier of the equation.
Therefore, we can naturally obtain the conservation law corresponding to each adjoint symmetry.
\begin{equation}
	\begin{aligned}
		&\phi_t^1=u_xv_x+vu_{xx}+uv_{xx}+\frac{3}{4}u^3v+\frac{9}{4}v^2u+\frac{3}{4}uu_x^2+\frac{3}{4}u^2u_{xx},\\
		&\phi_x^1=\frac{27}{8}u^2v^2+\frac{v_x^2}{2}+\frac{3}{4}v^3+\frac{u_{xx}^2}{6}-uv_{xt}+\frac{3}{4}u^4v+\frac{3}{8}u^2u_x^2+\frac{u^3u_{xx}}{4}+\frac{3uvu_{xx}}{2}+uu_xv_x-\frac{u_x^2v}{4}-\frac{3u^2u_{xt}}{4}.
	\end{aligned}
\end{equation}
for $Q_1$.
\begin{equation}
	\begin{aligned}
	&\phi_t^2=tuv-xv,\\
		&\phi_x^2=\frac{tv^2}{2}-xuv+tu^2v+\frac{t}{3}uu_{xx}-\frac{t}{6}u_x^2-\frac{xu_{xx}}{3}+\frac{u_x}{3},
	\end{aligned}
\end{equation}
for $Q_2$.
\begin{equation}
	\begin{aligned}
		&\phi_t^3=v^2u+\frac{u^3v}{2}+\frac{u^2u_{xx}}{6}+\frac{vu_{xx}}{3}+\frac{u_xu_t}{3},
		&\phi_x^3=\frac{v^2}{2}+\frac{u^2v}{2}-\frac{u_x^2}{6},
	\end{aligned}
\end{equation}
for $Q_3$.
\begin{equation}
	\begin{aligned}
			&\phi_t^4=uv,\\
		&\phi_x^4=\frac{v^2}{2}+vu^2+\frac{uu_{xx}}{3}-\frac{u_x^2}{6},
	\end{aligned}
\end{equation}
for $Q_4$.
\begin{equation}
	\begin{aligned}
		&\phi_t^5=u+v
		&\phi_x^5=\frac{u^2}{2}+v+uv+\frac{u_{xx}}{3},
	\end{aligned}
\end{equation}
for $Q_5+Q_6$.
\subsection{Symmetry actions and brackets for adjoint-symmetries}
In this section, we will use the symmetry of the target equation to act on its adjoint symmetry and obtain a Lie algebra of it.
First, we introduce some theorems.~\\
\textbf{Theorem 4:}
For any regular partial differential system, there are two actions of symmetries:
\begin{align}
	&Q_A\overset{X_p}{\longrightarrow}Q^{'}(P)_A+R_p^{*}(Q)_A,\\
	&Q_A\overset{X_p}{\longrightarrow}R_{p}^{*}(Q)_A-R_Q^{*}(P)_A,
\end{align}
on the linear space of adjoint-symmetries. The second symmetry action (35) maps adjoint-symmetries
into conservation law multipliers. The difference of the first and second actions yields the linear
mapping
\begin{equation}
	Q_A\overset{X_p}{\longrightarrow}Q^{'}(P)_A+R_Q^{*}(P)_A,
\end{equation}
where $R_p,R_Q$ mean the operator from the determining equation of the symmetry and adjoint sysmetry of the target equation $R_p(G)=G^{'}(p),R_Q(G)=G^{'^{*}}(Q)$ and $P$ and $Q$ is symmetry and adjoint symmetry of the PDE system.\par
The action (36) will be trivial when the adjoint-symmetry is a conservation law multiplier, as follows from the relation (36) which holds under certain mild conditions on the form of the PDE system
$G^A=0$ and the functions $Q_A$. ~\\
\textbf{Theorem 5:} Fix an adjoint-symmetry $Q_A$ in AdjSymmG, and let $S_Q$ be the dual linear operator $S_Q(P)_A:=S_p(Q)_A$ associated with a symmetry action $S_p$ on AdjSymmG. If the kernel of $S_Q$ is an ideal in SymmG,
then
\begin{equation}
	^{Q}\left[Q_1,Q_2\right]_A:=S_Q([S_Q^{-1}Q_1,S_Q^{-1}Q_2])_A,
\end{equation}
defines a bilinear bracket on the linear space $S_Q(Symm_G)\in AdjSymm_G.$ This bracket can be
expressed as
\begin{equation}
	^{Q}\left[Q_1,Q_2\right]_A=Q_2^{'}(S_Q^{-1}Q_1)-Q_1^{'}(S_Q^{-1}Q_2)-S_Q^{'}(S_Q^{-1}Q_2)(S_Q^{-1}Q_1)+S_Q^{'}(S_Q^{-1}Q_1)(S_Q^{-1}Q_2),
\end{equation}
where $S_Q^{'}$ denotes the Frechet derivative of $S_Q$.
In the previous part, we have solved a series of symmetries 
\begin{equation}
	\begin{aligned}
		&P_1=(-u_t,-v_t),P_2=(-u_x,-v_x),P_3=(1-tu_x,-tv_x),P_4=(-\frac{u}{2}-\frac{x}{2}u_x-tu_t,-v-\frac{x}{2}v_x-tv_t),
	\end{aligned}
\end{equation}
and adjoint symmetries of the target equation
\begin{equation}
	\begin{aligned}
		&Q_1=(v_{xx}+\frac{9}{4}v^2+\frac{9}{4}u^2v+\frac{6uu_{xx}+3u_x^2}{4},u_{xx}+\frac{3u^3}{4}+\frac{9}{2}uv),\\
		&Q_2=(tv,tu-x),Q_3=(tv,\frac{u^2}{2}+v),Q_4=(v,u),Q_5=(1,0),Q_6=(0,1).
	\end{aligned}
\end{equation}
Their algebra has the non-zero commutators
\begin{equation}
	\left[P_1,P_3\right]=P_4,\left[P_1,P_4\right]=P_1,\left[P_2,P_4\right]=\frac{P_2}{2},\left[P_3,P_4\right]=-\frac{P_3}{2}.
\end{equation}
Since the above six adjoint equations are all multipliers of the target equation, action (36) is trivial for every $Q_i$.
And We can find that a is an evolution equation system, so  we have a conclusion $Q^{'}=-R_p^{*}$ and action (34) is the same as action (35). Then we choose action (34) to analysis.
We can get $R_{p_i}$ by caluation
\begin{equation}
	\begin{aligned}
		&R_{p_1}=\left(\begin{array}{cc}
			-D_t & 0\\
			0 & -D_t \\
		\end{array}\right),
	R_{p_2}=\left(\begin{array}{cc}
	-D_x & 0\\
	0 & -D_x \\
\end{array}\right),\\
	&R_{p_3}=\left(\begin{array}{cc}
	-tD_x & 0\\
	0 & -tD_x \\
\end{array}\right),
	R_{p_4}=\left(\begin{array}{cc}
	-\frac{3}{2}-\frac{x}{2}D_x-tD_t & 0\\
	0 & -\frac{x}{2}D_x-tD_t-2 \\
\end{array}\right).
\end{aligned}
\end{equation}
\begin{table}{}
	\begin{center}
	\caption{ Symmetry action (34) on
		adjoint-symmetries}
	\begin{tabular}{ccccc}
		\hline
		  & $P_1$ & $P_2$ & $P_3$ & $P_4$\\
		\hline
		$Q_1$ & 0 & 0 & $\frac{9}{2}Q_3$ & $-2Q_1$\\
		$Q_2$ & $Q_4$ & $-Q_6$ & 0 & 0\\
			$Q_3$ & 0 & 0 & $Q_4$ & $-\frac{3}{2}Q_3$\\
				$Q_4$ & 0 & 0 & $Q_6$ & $-Q_4$\\
					$Q_5$ & 0 & 0 & 0 & 0\\
						$Q_6$ & 0 & 0 & 0 & 0\\
		\hline
	\end{tabular}
	\end{center}
\end{table}
~\\
According to Theorem 5, we can know that the condition for the existence of bracket is kernel of $S_Q$ is an ideal for an fixed adjoint symmetry. Therefore, after verification, we get $Q_i$ satisfies the condition when $i=1,3,\cdots,6$. However, the resulting bracket is trivial when $i=5,6$. For example, when we fix $Q_1$, we get $ker(S_{Q_1})=\left\{0,P_1,P_2\right\}$. 
For any $P_{j}\in Symm_G, \left[P_i,P_j\right]\in ker(S_{Q_1})$, where $P_i\in ker(S_{Q_1})$.
Therefore, we can find the following commutators after the action of the adjoint symmetry by the commutators between the symmetries from table 1:
\begin{equation}
	\begin{aligned}
		&^{Q_1}\left[Q_1,Q_3\right]=S_{Q_1}\left[S_{Q_1}^{-1}Q_1,S_{Q_1}^{-1}Q_3\right]=S_{Q_1}\left[-\frac{1}{2}P_4,\frac{2}{9}P_3\right]=S_{Q_1}(-\frac{P_3}{18})=-\frac{Q_3}{4},\\
		&^{Q_3}\left[Q_3,Q_4\right]=S_{Q_3}\left[S_{Q_3}^{-1}Q_3,S_{Q_3}^{-1}Q_4\right]=S_{Q_3}\left[-\frac{2}{3}P_4,P_3\right]=S_{Q_1}(\frac{P_3}{3})=\frac{Q_4}{3},\\
		&^{Q_4}\left[Q_4,Q_6\right]=S_{Q_4}\left[S_{Q_4}^{-1}Q_4,S_{Q_4}^{-1}Q_6\right]=S_{Q_4}\left[-P_4,P_3\right]=S_{Q_4}(\frac{P_3}{2})=\frac{Q_6}{2}.
	\end{aligned}
\end{equation}
\section{Compute conservation laws by Noether's theorem}
In this part, we try to use Noether's theorem to solve the conservation laws of the (1+1)-dimensional dispersive long-wave equation.
First, we introduce the operator 
$$E_{\gamma}=\frac{\partial}{\partial u^{\gamma}}-D_{i}\frac{\partial}{\partial u_{i}^{\gamma}}+D_{i}D_{j}\frac{\partial}{\partial u_{ij}^{\gamma}}+\cdots+(-1)^{k}D_{i_{1}}D_{i_{2}}\cdots D_{i_{k}}\frac{\partial}{\partial u_{i_{1}i_{2}\cdots i_{k}}^{\gamma}}.$$
Then let
\begin{equation}
	\begin{aligned}
		W^{i}[u,v]=&v^{\gamma}[\frac{\partial L}{\partial u_{i}^{\gamma}}+\cdots+(-1)^{k-1}D_{i_{1}}D_{i_{2}}\cdots D_{i_{k-1}}\frac{\partial L}{\partial u_{ii_{1}i_{2}\cdots i_{k-1}}^{\gamma}}]+(D_{i_{1}}v^{\gamma})[\frac{\partial L}{\partial u_{i_{1}i}^{\gamma}}+\cdots \\
		&+(-1)^{k-2}D_{i_{2}}D_{i_{3}}\cdots D_{i_{k-1}}\frac{\partial L}{\partial u_{ii_{1}i_{2}\cdots i_{k-1}}^{\gamma}}]+\cdots+(D_{i_{1}}D_{i_{2}}\cdots D_{i_{k-1}}v^{\gamma})\frac{\partial L}{\partial u_{ii_{1}i_{2}\cdots i_{k-1}}}.
	\end{aligned}
\end{equation}
~\\
\textbf{Definition 3}
A transformation $x^{*}=x,u^{*}=u+\epsilon \eta(x,u,u_{(1)},\dots,u_{(p)})$ is a variation symmetry of the action intergral $J[u]$ if for any $u(x)$ there exists some vector function 
\begin{equation}
	A(x,u,u_{(1)},\dots,u_{(r)})=(A^{1},A^{2},\dots,A^{n}),  \notag
\end{equation}
of $x,u$ and its derivatives to some finite order $r$, such that
\begin{equation}
	U^{k}L=D_{i}A^{i},
\end{equation}
where $U$ represents the symmetric infinitesimal operator of the target equation and $L$ represents Euler-Lagrangian equation of the equation, i.e. $E_{r}L=G$.
~\\
\textbf{Theorem 4} Let $U=\eta^{\gamma}\frac{\partial}{\partial u^{\gamma}}$ be the infinitesimal generator of transformations  $x^{*}=x,u^{*}=u+\epsilon \eta(x,u,u_{(1)},\dots,u_{(p)})$ and let $U^{(k)}$  be its $k$th extension. If $U$ is the infinitesimal generator of a variational symmetry of an action integral so that $U^{(k)}=D_{i}A^{i}$ holds for any $u(x)$, then the conservation law
\begin{equation}
	D_{i}(W^{i}[u,\eta]-A^{i})=0,
\end{equation}
holds for any solution $u(\boldsymbol{x})$ of the Euler-Lagrange equations $E_{\gamma}(L)=0, \gamma=1,2,\dots,m$.
Then for the (1+1) dimension dispersive long-wave equation
\begin{equation}
	\begin{aligned}
		&u_t+uu_x+v_x=0,\\
		&v_t+u_xv+v_xu+\frac{1}{3}u_{xxx}=0.
	\end{aligned}
\end{equation}
It has no Lagrangian $L$. However if we make the 
substitution $q=u_{x},r=v_x$ in (47) then the Eq. (47) becomes
\begin{equation}
	\begin{aligned}
		&q_{xt}+q_xq_{xx}+r_{xx}=0,\\
		&r_{xt}+q_{xx}r_{x}+q_{x}r_{xx}+\frac{1}{3}q_{xxxx}=0.
	\end{aligned}
\end{equation}
is the Euler-Lagrange equation for
\begin{equation}
	L=-q_{x}r_{t}-\frac{r_{x}^2}{2}-\frac{q_x^2r_x}{2}+\frac{1}{6}q_{xx}^2.
\end{equation}
Eq. (47) admits an infinite sequence of Lie-Backlund 
symmetries. The first few infinitesimal generators are:
\begin{equation}
	\begin{aligned}
		&U_{1}=u_{x}\frac{\partial}{\partial u}+v_x\frac{\partial}{\partial v}, U_{2}=u_{t}\frac{\partial}{\partial u}+v_{t}\frac{\partial}{\partial v}, U_{3}=(1-tu_{x})\frac{\partial}{\partial u}-tv_x\partial v,\\
		&U_{4}=(\frac{u}{2}+\frac{xu_x}{2}+tu_t)\frac{\partial}{\partial u}+(v+\frac{x}{2}v_x+tv_t)\frac{\partial}{\partial v}.
	\end{aligned}
\end{equation}
Since $q=\int udx,r=\int vdx$,  the corresponding infinitesimal generators for (are obtained by integrating with respect to $x$) 
the coefficients of the infinitesimal generators (50) and then replacing $u$ and $v$
by $q_x$ and $r_{x}$:
\begin{equation}
	\begin{aligned}
		&V_{1}=q_{x}\frac{\partial}{\partial q}+r_x\frac{\partial}{\partial r}, V_{2}=q_{t}\frac{\partial}{\partial q}+r_t\frac{\partial}{\partial r}, V_{3}=(x-tq_x)\frac{\partial}{\partial q}-tr_x\frac{\partial}{\partial r},\\
		&V_{4}=(tq_t+\frac{x}{2}q_x)\frac{\partial}{\partial q}+(\frac{r}{2}+tr_t+\frac{x}{2}r_x)\frac{\partial}{\partial r}.
	\end{aligned}
\end{equation}
Firstly, we calculate $W^{i}[u,\eta]$ by the formula 
\begin{equation}
	V^{(k)}L=E_{\gamma}(L)\eta+D_{i}W^{i}.
\end{equation}
We use the general formula $V=\eta_1 \frac{\partial}{\partial q}+\eta_2 \frac{\partial}{\partial r}$.
Then the Eq. (52) becomes 
\begin{equation}
	\begin{aligned}
		D_iW^i=&U^{(k)}L-E_{\gamma}(L)\eta=-r_tD_x\eta_1-q_xD_t\eta_2-r_xD_x\eta_2-\frac{q_x^2}{2}D_x\eta_2-q_xr_xD_x\eta_1+\frac{1}{3}q_{xx}D_x^2\eta_1\\&-\eta_1(r_{xt}+q_{xx}r_{x}+q_xr_{xx}+\frac{1}{3}q_{xxxx})-\eta_2(q_{xt}+q_xq_{xx}+r_{xx})\\
		=&D_x(-\eta_1r_t-r_x\eta_2-\frac{q_x^2\eta_2}{2}-q_xr_x\eta_1-\frac{\eta_1q_{xxx}}{3}+\frac{q_{xx}}{3}D_x\eta_1)+D_t(-q_x\eta_2).
	\end{aligned}
\end{equation}
Therefore we can obtain $W^1=-\eta_1r_t-r_x\eta_2-\frac{q_x^2\eta_2}{2}-q_xr_x\eta_1-\frac{\eta_1q_{xxx}}{3}+\frac{q_{xx}}{3}D_x\eta_1,W^2=-q_x\eta_2.$Then the conservation laws $D_{x}f^{1}+D_{y}f^{2}=0$ where $f^{1}=W^{1}-A^{1},f^{2}=W^{2}-A^{2}$ are obtained as follows:
\par(i) From $V_{1}$:$V_{1}L=D_{x}L$. Hence $A^{1}=L,A^{2}=0$;
\begin{equation}
	\begin{aligned}
		&f^1_1=-\frac{r_x^2}{2}-q_x^2r_x-\frac{q_xq_{xxx}}{3}+\frac{q_{xx}^2}{6},\\
		&f^2_1=-q_xr_x.
	\end{aligned}
\end{equation}
We can simply verify it:
$$D_xf^1_1+D_tf^2_1=-r_xr_{xx}-2q_xr_xq_{xx}-q_x^2r_{xx}-\frac{q_xq_{xxxx}}{3}-q_{xt}r_x-q_xr_{xt}=0,$$
where $(q,r)$ is a solution of Eq. (48).
\par(ii) From $V_{2}$:$V_{2}L=D_{t}L$. Hence $A^{1}=0,A^{2}=L$;
\begin{equation}
	\begin{aligned}
		&f^1_2=-q_tr_t-r_tr_x-\frac{q_x^2r_t}{2}-q_tq_xr_x-\frac{q_tq_{xxx}}{3}+\frac{q_{xx}q_{xt}}{3},\\
		&f^2_2=\frac{r_{x}^2}{2}+\frac{q_x^2r_x}{2}-\frac{q_{xx}^2}{6}.
	\end{aligned}
\end{equation}
We can simply verify it:
$$D_xf^1_2+D_tf^2_2=-q_{xt}r_t-q_tr_{xt}-r_tr_{xx}-r_tq_xq_{xx}-q_tq_{xx}r_x-q_tq_xr_{xx}-\frac{q_tq_{xxxx}}{3}=0,$$
where $(q,r)$ is a solution of Eq. (48).
\par(iii) From $V_{3}$:$V_{3}L=D_{x}(tq_xr_t+\frac{t}{2}r_x^2+\frac{tq_x^2r_x}{2}-\frac{t}{6}q_{xx}^2)+D_t(-r)$. Hence $A^{1}=tq_xr_t+\frac{t}{2}r_x^2+\frac{tq_x^2r_x}{2}-\frac{t}{6}q_{xx}^2$,$A^{2}=-r$;
\begin{equation}
	\begin{aligned}
		&f^1_3=-xr_t+\frac{tr_x^2}{2}-q_xr_xx+tq_x^2r_x-\frac{xq_{xxx}}{3}+\frac{tq_xq_{xxx}}{3}+\frac{q_{xx}}{3}-\frac{tq_{xx}^2}{6},\\
		&f^2_3=tq_xr_x+r.
	\end{aligned}
\end{equation}
We can simply verify it:
\begin{equation}
	\begin{aligned}
		D_xf^1_3+D_tf^2_3=&-xr_{xt}+tq_xr_{xt}+tr_xr_{xx}+2tq_xr_xq_{xx}-q_{xx}r_{x}x-q_xr_{xx}x+tq_x^2r_{xx}\\&-\frac{x}{3}q_{xxxx}+\frac{tq_{x}q_{xxxx}}{3}+q_{xt}tr_x=0.\notag
	\end{aligned}
\end{equation}
where $(q,r)$ is a solution of Eq. (48).
\par
(iv) From $V_{4}$: After using $V_4$ to act on $L$, we find that it can not be reduced to a conserved form. So $V_4$ is not the variation symmetry.
\section{Solving conservation law by Ibragimov method}
In this part we try to solve the equation by Ibragimov method to obtain the conservation law. First, we introduce a theorem:
~\\ \textbf{Theorem 1.} Any Lie point, Lie-B{\"a}cklund, and nonlocal symmetry
$$X=\xi^{i}(x,u,u_{(1)},\ldots)\frac{\partial}{\partial x^{i}}+\eta^{\alpha}(x,u,u_{(1)},\ldots)\frac{\partial}{\partial u^{\alpha}}$$ leads to the conservation law $D_{i}(C^{i})=0$, i.e.
\begin{equation}
	\begin{aligned}
		C^{i} =&\xi^{i}\mathscr{L}+W^{\alpha}[\frac{\partial \mathscr{L}}{\partial u^{\alpha}_{i}}-D_{j}(\frac{\partial \mathscr{L}}{\partial u^{\alpha}_{ij}})+D_{j}D_{j}(\frac{\partial \mathscr{L}}{\partial u^{\alpha}_{ijk}})-\cdots]\\
		& +D_{j}(W^{\alpha})[\frac{\partial \mathscr{L}}{\partial u^{\alpha}_{ij}}-D_{k}(\frac{\partial \mathscr{L}}{\partial u^{\alpha}_{ijk}})+\cdots]  +D_{j}D_{k}(W^{\alpha})[\frac{\partial \mathscr{L}}{\partial u^{\alpha}_{ijk}}-\cdots]+\cdots,
	\end{aligned} 
\end{equation}
where $$
\mathscr{L}=\sum\limits_{i=1}^{m}v^{i}F_{i}(x,u,u_{(1)},\ldots,u_{(s)}), W^{\alpha}=\eta^{\alpha}-\xi^{j}u_{j}^{\alpha},\alpha=1,...,m.
$$
Then we start with a system of $m$ differential equations with the following forms
\begin{equation}
	F_{\alpha}(x,u,u_{(1)},...,u_{(s)})=0,\alpha=1,2,\ldots,m, 
\end{equation}
which admits the adjoint equations
\begin{equation}
	F_{\alpha}^{*}(x,u^{1},u^{2},u^{1}_{(1)},u^{2}_{(1)},\ldots,u^{1}_{(s)},u^{2}_{(s)})=0,\alpha=1,2,\ldots,m,
\end{equation}
where
\begin{equation}
	\begin{aligned}
		&F_{\alpha}^{*}(x,u^{1},u^{2},u^{1}_{(1)},u^{2}_{(1)},\ldots,u^{1}_{(s)},u^{2}_{(s)})=\frac{\delta\mathcal{L}}{\delta u^{\alpha}},\\
		&u^{1}_{(1)}={u^{\alpha}_{i}},u^{1}_{(s)}={u^{\alpha}_{i_{1}\cdots i_{s}}},
		u^{2}_{(1)}={v^{\alpha}_{i}},u^{2}_{(s)}={v^{\alpha}_{i_{1}\cdots i_{s}}},
	\end{aligned}
\end{equation}
where $\mathcal{L}$ means the formal Lagrangian for Eq. (98), i.e.
\begin{equation}
	\mathcal{L}=\sum\limits_{i=1}^{m}v^{i}F_{i}(x,u,u_{(1)},\ldots,u_{(s)}), 
\end{equation}
$\frac{\delta}{\delta u^{\alpha}}$ means the variational derivative with the following forms
$$\frac{\delta}{\delta u^{\alpha}}=\frac{\partial}{\partial u^{\alpha}}+\sum\limits^{\infty}_{s=1}(-1)^{s}D_{i_{i}}\cdots D_{i_{s}}\frac{\partial}{\partial u^{\alpha}_{i_{1}\cdots i_{s}}}.$$
\textbf{Definition 2.} We say that the differential Eq.(58) is strictly self-adjoint
if the adjoint Eq.(59) becomes equivalent to the original Eq.(60)
upon the substitution
$$v^{\alpha}=u^{\alpha},\alpha=1,2,\ldots,m.$$
It means that the equation
\begin{equation}
	F_{\alpha}^{*}(x,u,v,u_{(1)},v_{(1)},\ldots,u_{(s)},v_{(s)})=F_{\alpha}(x,u,v,u_{(1)},v_{(1)},\ldots,u_{(s)},v_{(s)}). 
\end{equation}
Next we begin to study the Eq. (1). We note $\mathcal{L}=w^1(v_t+u_xv+uv_x+\frac{1}{3}u_{xxx})+w^2(u_{t}+uu_{x}+v_x)$, where $w^1,w^2$ have the expressions $w^1(x,t),w^2(x,t)$.
~\\
\textbf{Theorem 2.} The (1+1)-dimensional dispersive long-wave equation
\begin{equation}
	\begin{aligned}
		&v_t+u_xv+uv_x+\frac{1}{3}u_{xxx}=0,\\
			&u_t+uu_x+v_x=0.
	\end{aligned}
\end{equation}
is strictly self-adjoint.
\par \noindent Proof: The formal Lagrangian of the (1+1)-dimensional dispersive long-wave equation reads
\begin{equation}
\mathcal{L}=w^1(v_t+u_xv+uv_x+\frac{1}{3}u_{xxx})+w^2(u_{t}+uu_{x}+v_x),
\end{equation}
Then we can get
\begin{equation}
	F^*=
	\left(\begin{array}{c}
		\frac{\delta \mathcal{L}}{\delta u}  \\
		\frac{\delta \mathcal{L}}{\delta v}    \\
	\end{array}\right)
=	\left(\begin{array}{c}
-vD_xw^1-\frac{1}{3}D_x^3w^1-D_tw^2-uD_xw^2 \\
-vD_xw^1-uD_xw^1-D_xw^2   \\
\end{array}\right),
\end{equation}
If we make $w^1=u,w^2=v$,
\begin{equation}
\begin{cases} 
	\frac{\delta \mathcal{L}}{\delta u}=-v_t-vu_x-uv_x-\frac{1}{3}u_{xxx}\\ 
\frac{\delta \mathcal{L}}{\delta v}=-u_t-uu_x-v_x
\end{cases} 
=-F.
\end{equation}
Therefore, according to the Definition 2, the Eq. (1) is nonlinear self-adjoint.
~\\
Next, we use Theorem 1 and the symmetry of the Eq. (1) to solve the corresponding conservation law.
It has symmetries:
\begin{equation}
	\begin{aligned}
		&X_{1}=\frac{\partial}{\partial t},X_{2}=\frac{\partial}{\partial x},X_{3}=t\frac{\partial}{\partial x}+\frac{\partial}{\partial u},\\
		&X_{4}=\frac{x}{2}\frac{\partial}{\partial x}+t\frac{\partial}{\partial t}-\frac{u}{2}\frac{\partial}{\partial u}-v\frac{\partial}{\partial v}.\\
	\end{aligned}
\end{equation}
According to the formula (57), we have
$$W^1_1=-u_x,W^1_2=-v_x,$$ for $X_1$.
Then we obtain 
\begin{equation}
	\begin{aligned}
		C_x^1=&\xi^x\mathcal{L}+W^1_1[\frac{\partial \mathcal{L}}{\partial u_x}+D_x^2(\frac{\partial \mathcal{L}}{\partial u_{xxx}})]+W^1_2\frac{\partial \mathcal{L}}{\partial v_x}+D_xW^1_1(-D_x)(\frac{\partial \mathcal{L}}{\partial u_{xxx}})+D_x^2W^1_1\frac{\partial \mathcal{L}}{\partial u_{xxx}}\\
		=&w^1v_t+w^2u_t-\frac{u_xw^1_{xx}}{3}+\frac{w^1_xu_{xx}}{3},\\
		C_t^1=&\xi^tL+W^1\frac{\partial \mathcal{L}}{\partial u_t}+W^2\frac{\partial \mathcal{L}}{\partial v_t}\\
		=&-u_xw^2-v_xw^1.
	\end{aligned}
\end{equation}
Then we have $W^2_1=-u_t,W^2_2=-v_t$ for $X_2$, and the corresponding conservation law is
\begin{equation}
	\begin{aligned}
		&C_x^2=-u_tvw^1-uu_tw^2-\frac{u_tw^1_{xx}}{3}-uv_tw^1-v_tw^2+\frac{u_{xt}w^1_x}{3}-\frac{u_{xxt}w^1}{3},\\
		&C_t^2=w^1u_xv+w^1v_xu+\frac{w^1u_{xxx}}{3}+uw^2u_x+w^2v_x.
	\end{aligned}
\end{equation}
For $X_3$, we get $W^3_1=1-tu_x,W^3_2=-tv_x$ and the conservation law is
\begin{equation}
	\begin{aligned}
		&C_x^3=tw^1v_t+tw^2u_t+vw^1+uw^2+\frac{w^1_{xx}}{3}-\frac{tu_xw^1_{xx}}{3}+\frac{tu_{xx}w^1_x}{3},\\
		&C_t^3=w^2-w^2tu_x-tv_xw^1.
	\end{aligned}
\end{equation}
For $X_4$, we have $W^4_1=-\frac{u}{2}-\frac{xu_x}{2}-tu_t,W^4_2=-v-\frac{xv_x}{2}-tv_t$. Then the corresponding conservation law is 
 \begin{equation}
 	\begin{aligned}
 		C_x^4=&-tvu_tw^1-tu_tuw^2-tv_tuw^1-tv_tw^2+\frac{w^1_xxu_{xx}}{6}+\frac{w^1_xtu_{xt}}{3}-\frac{w^1tu_{xxt}}{3}+\frac{xw^1v_t}{2}+\frac{xw^2u_t}{2}-\frac{3uvw^1}{2}\\&-\frac{xu_xw^1_{xx}}{6}-\frac{tu_tw^1_{xx}}{3}-\frac{w^2u^2}{2}-\frac{uw^1_{xx}}{6}-vw^2+\frac{w^1_xu_x}{3}-\frac{w^1u_{xx}}{2},\\
 		C_t^4=&tw^1u_xv+tw^1v_xu+\frac{tw^1}{3}u_{xxx}+tw^2uu_x+tw^2v_x-\frac{uw^2}{2}-\frac{w^2xu_x}{2}-vw^1-\frac{w^1xv_x}{2}.
 	\end{aligned}
 \end{equation}
\section{Hamiltonian Structure and Line Soliton Solution}
For the (1+1) dimensional dispersive long-wave equation equation
\begin{equation}
	\begin{aligned}
		&u_t+uu_x+v_x=0,\\
		&v_t+u_xv+uv_x+\frac{1}{3}u_{xxx}=0.
	\end{aligned}
\end{equation}
We notice that it has a Hamiltonian formulation 
$u_{t}=-\mathcal{D}(\frac{\delta H}{\delta u})$
\begin{equation}
	\left(\begin{array}{c}
	u \\
		v   \\
	\end{array}\right)_t=-{\mathcal{D}}\left(\begin{array}{c}
	\frac{\delta H }{\delta u}\\
	\frac{\delta H}{\delta v}   \\
\end{array}\right)=-\mathcal{D}\left(\begin{array}{c}
uv+\frac{u_{xx}}{3}\\
v+\frac{u^2}{2}  \\
\end{array}\right),
\end{equation}
where $H=\int \frac{v^2}{2}+\frac{u^2v}{2}-\frac{u_x^2}{6}dx$  is the Hamiltonian functional, and $\mathcal{D}=\left(\begin{array}{cc}
	0 & D_x \\
	D_x & 0
\end{array}\right)$ is a Hamiltonian operator. Then since $\mathcal{D}$ is a Hamiltonian operator, it can map adjoint-symmetries into symmetries, so $\mathcal{J}=\left(\begin{array}{cc}
0 & D_x^{-1} \\
D_x^{-1} & 0
\end{array}\right)$  can map symmetries into adjoint-symmetries. And we can use the above symmetry to get the adjoint symmetry of some objective equations. Applying this latter
operator to the scaling symmetry, we obtain the adjoint-symmetry:
\begin{equation}
	\begin{aligned}
		&Q_{1}=\mathcal{J}\left(\begin{array}{c}
		-u_t \\
		-v_t
	\end{array}\right)=\left(\begin{array}{c}
	-r_t \\
	-q_t
\end{array}\right),
Q_{2}=\mathcal{J}\left(\begin{array}{c}
	-u_x \\
	-v_x
\end{array}\right)=\left(\begin{array}{c}
	v \\
	u
\end{array}\right),\\&Q_{3}=\mathcal{J}\left(\begin{array}{c}
1-tu_x \\
-tv_x
\end{array}\right)=\left(\begin{array}{c}
-tv \\
x-tu
\end{array}\right),
		Q_{4}=\mathcal{J}\left(\begin{array}{c}
			-\frac{u}{2}-\frac{xu_x}{2}-tu_t \\
			-v-\frac{xv_x}{2}-tv_t
		\end{array}\right)=\left(\begin{array}{c}
			-r-tr_t-\frac{xv}{2}+r \\
			-\frac{q}{2}-\frac{xu}{2}+q-tq_t
		\end{array}\right),
	\end{aligned}
\end{equation}
where $u=q_{x},v=r_x$.~\\
Next, we try to solve the soliton solution by using the conservation law constructed by the adjoint symmetry.
A line soliton is a solitary traveling wave $u=U(x-\mu t),v=V(x-\mu t)$ in one dimension where the parameter $\mu$ means the speed of the wave. Then we study the conservation laws of the ZK equation $\phi^{t},\ \phi^{x}$ which doesn't contain the variables $t,x.$
Then the conservation law is obtained by reduction
\begin{equation}
	D_{t}|_{u=U(\xi)}=-\mu \frac{d}{d \xi}, D_{x}|_{u=U(\xi)}=\frac{d}{d \xi},\xi=x-\mu t,
\end{equation}
yielding
\begin{equation}
	\frac{d}{d\xi}(\phi^{x}-\mu \phi^{t})=0.
\end{equation}
So $(\phi^{x}-\mu \phi^{t})=C.$
Then we begin with the equation (1). Using the transformation $u(x,t)=U(\xi),v(x,t)=V(\xi),$
we can obtain the nonlinear ordinary differential equation:
\begin{equation}
	\begin{aligned}
		&-\mu U^{'}+UU^{'}+V^{'}=0,\\
		&-\mu V^{'}+UV^{'}+VU^{'}+\frac{U^{'''}}{3}=0.
	\end{aligned}
\end{equation}
for $U(\xi),V(\xi)$.
Conservation laws (29), (31), (32) and (33) do not contain the variables $t,x$. When the first integral formula $(\phi^{x}-\mu \phi^{t})=C$ is applied to these three conservation laws, we obtain 
\begin{align}
	C_1=&\frac{27U^2V^2}{8}+\frac{(V^{'})^2}{2}+\frac{3V^3}{4}+\frac{(U^{''})^2}{6}+\mu UV^{''}+\mu \notag VU^{''}+\frac{3U^4V}{4}+\frac{3U^2(U^{'})^2}{8}+\frac{U^3U^{''}}{4}\\&+\frac{3UVU^{''}}{2}+UU^{'}V^{'}-\frac{V(U^{'})^2)}{4}+\frac{3\mu U^2U^{''}}{4}-\mu(U^{'}V^{'}+VU^{''}+V^{''}U+\frac{3U^3V}{4}+\frac{9V^2U}{4} 
	\notag\\&+\frac{3U(U^{'})^2}{4}+\frac{3U^2U^{''}}{4}) \\
	=&\frac{3U^4V}{4}-\frac{3U^3V\mu}{4}+\frac{U^3U^{''}}{4}+\frac{27U^2V^2}{8}+\frac{3U^2(U^{'})^2}{8}-\frac{9UV^2\mu}{4}-\frac{3U(U^{'})^2\mu)}{4}+\frac{3UVU^{''}}{2}\notag\\&+UU^{'}V^{'}+\frac{3V^3}{4}-\frac{V(U^{'})^2}{4}-U^{'}V^{'}\mu+\frac{(U^{''})^2}{6}+\frac{(V^{'})^2}{2}=0,\notag \\
	C_2=&\frac{U^3V}{2}-\frac{U^2V\mu}{2}+\frac{U^2U^{''}}{6}+V^2U-\frac{V^2\mu}{2}-\frac{\mu (U^{'})^2}{6}+\frac{VU^{''}}{3}=0,\\
	C_3=&\frac{V^2}{2}+U^{2}V+\frac{UU^{''}}{3}-\frac{(U^{'})^2}{6}-\mu UV=0, \\
    C_4=&UV+\frac{U^{''}}{3}+\frac{U^2}{2}+V-\mu U-\mu V=0.
\end{align}
We impose the asymptotic conditions $U,U^{'},U^{''},U^{'''}\rightarrow 0$ as $|\xi|\rightarrow \infty.$ Then we combine the formulas (78), (79), (80) and (81), then we can calculate its general line soliton solutions:
\begin{align}
	U_1=&\frac{2e^{C_1\mu\sqrt{3}}\mu}{e^{\mu\xi\sqrt{3}}(-1+\frac{e^{C_1\mu \sqrt{3}}}{e^{\mu \xi\sqrt{3}}})},
	V_1=-\frac{2e^{\mu \xi\sqrt{3}}\mu^2e^{C_1\mu\sqrt{3}}}{(e^{C_1\mu \sqrt{3}}-e^{\mu\xi\sqrt{3}})^2},\\
		U_2=&\frac{2e^{\xi\mu\sqrt{3}}\mu}{e^{C_1\mu\sqrt{3}}(-1+\frac{e^{\xi\mu \sqrt{3}}}{e^{\mu C_1\sqrt{3}}})},
	V_2=-\frac{2e^{\mu \xi\sqrt{3}}\mu^2e^{C_1\mu\sqrt{3}}}{(e^{C_1\mu \sqrt{3}}-e^{\mu\xi\sqrt{3}})^2}.
\end{align}
where $C_{1}$ is an arbitrary constant. 
\begin{figure}[htbp]
	\centering
	\subfigure{}{
		\begin{minipage}[t]{0.24\linewidth}
			\centering
			\includegraphics[width=1.2in]{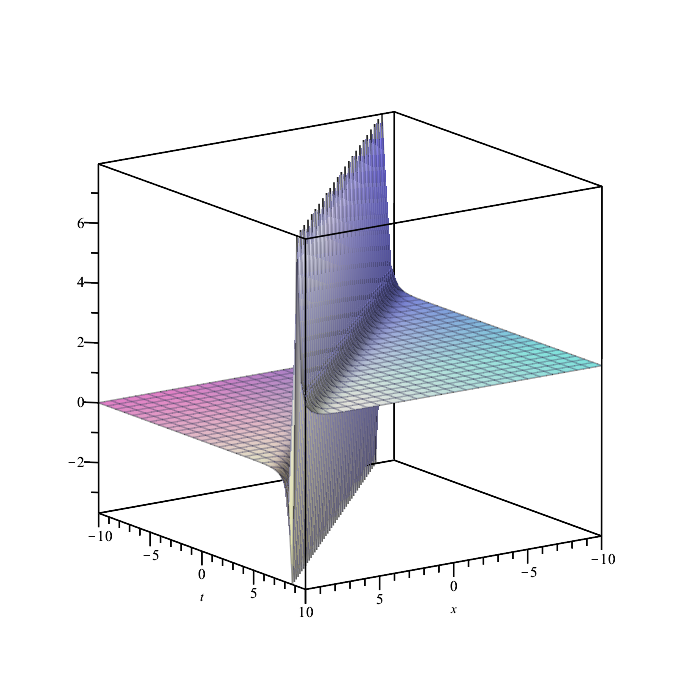}
			%\caption{fig 1}
		\end{minipage}%
	}%
	\subfigure{}{
		\begin{minipage}[t]{0.24\linewidth}
			\centering
			\includegraphics[width=1in]{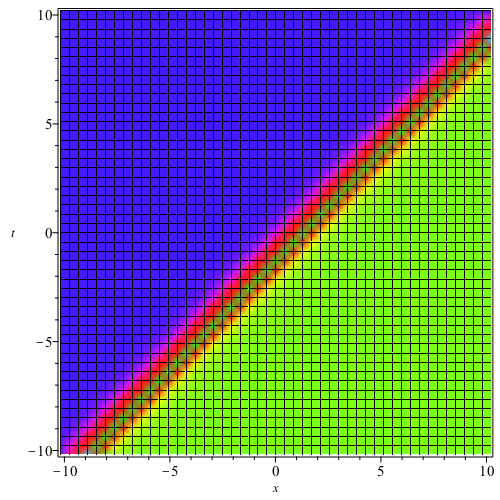}
			%\caption{fig 1}
		\end{minipage}%
	}%
	\subfigure{}{
		\begin{minipage}[t]{0.24\linewidth}
			\centering
			\includegraphics[width=1.2in]{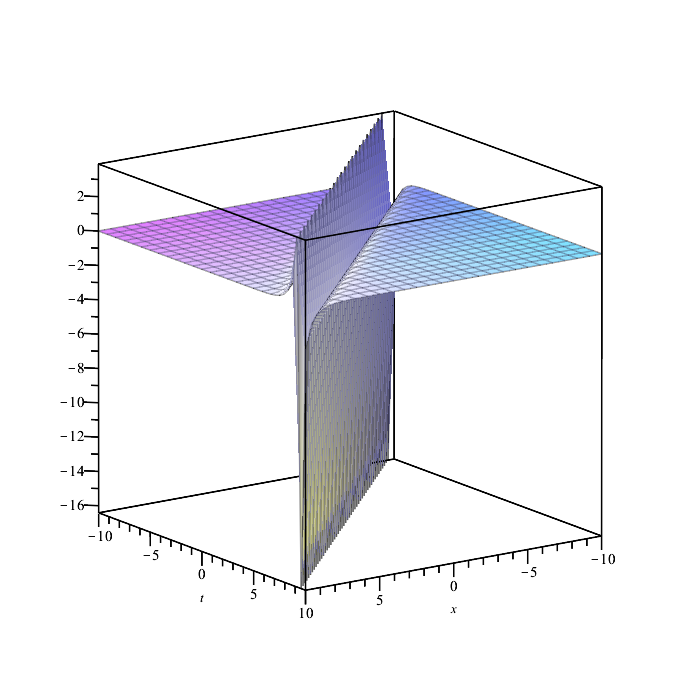}
			%\caption{fig 1}
		\end{minipage}%
	}%
	\subfigure{}{
		\begin{minipage}[t]{0.24\linewidth}
			\centering
			\includegraphics[width=1in]{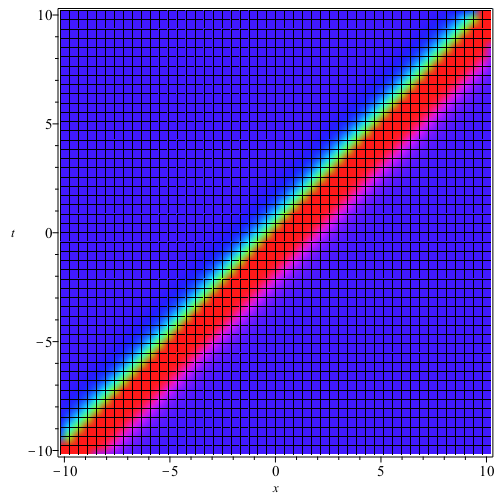}
			%\caption{fig 1}
		\end{minipage}%
	}%
	\centering
	\caption{3D plots and density plots of the $u(x,y,t)$ given by Eq. (138) for parameters $a_1 = 1, t = 1$ and  $a_1 = -1, t = 1$.}
	\label{Fig1}
\end{figure}
\begin{figure}[htbp]
	\centering
	\subfigure{}{
		\begin{minipage}[t]{0.24\linewidth}
			\centering
			\includegraphics[width=1.2in]{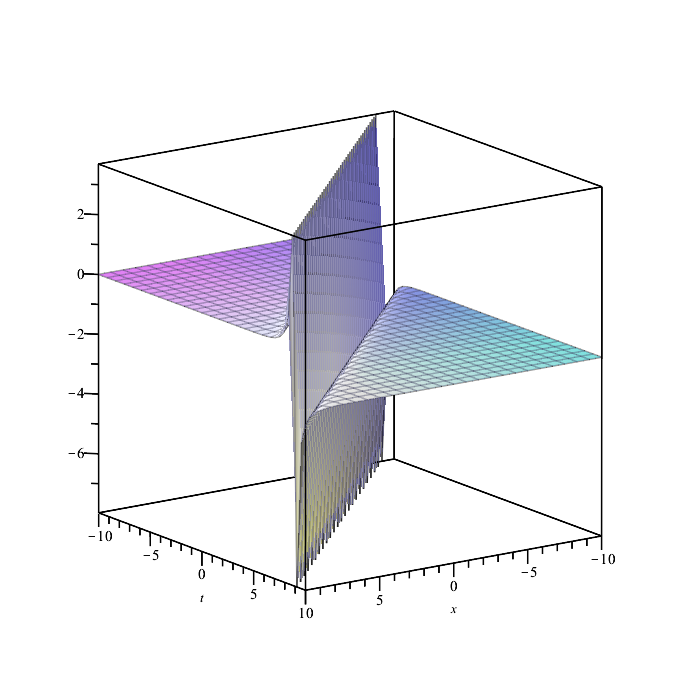}
			%\caption{fig 1}
		\end{minipage}%
	}%
	\subfigure{}{
		\begin{minipage}[t]{0.24\linewidth}
			\centering
			\includegraphics[width=1in]{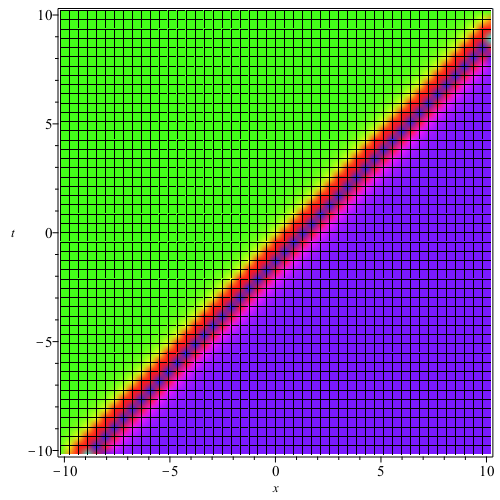}
			%\caption{fig 1}
		\end{minipage}%
	}%
	\subfigure{}{
		\begin{minipage}[t]{0.24\linewidth}
			\centering
			\includegraphics[width=1.2in]{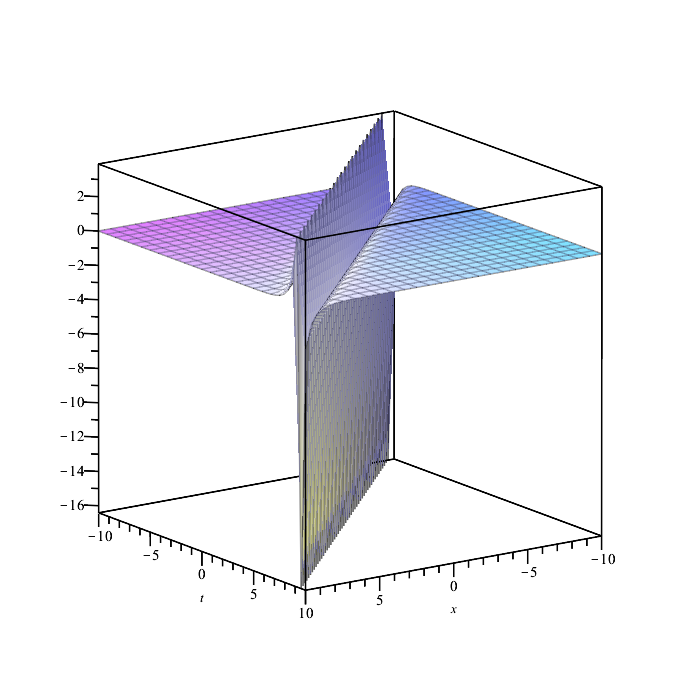}
			%\caption{fig 1}
		\end{minipage}%
	}%
	\subfigure{}{
		\begin{minipage}[t]{0.24\linewidth}
			\centering
			\includegraphics[width=1in]{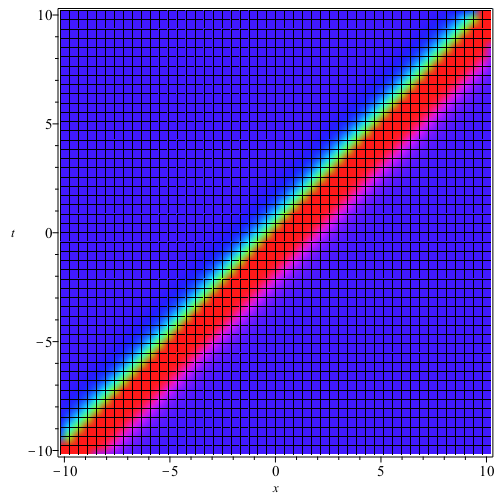}
			%\caption{fig 1}
		\end{minipage}%
	}%
	\centering
	\caption{3D plots and density plots of the $u(x,y,t)$ given by Eq. (138) for parameters $a_1 = 1, t = 1$ and  $a_1 = -1, t = 1$.}
	\label{Fig1}
\end{figure}
\section{Exact solutions of (1+1) dimensional dispersive long-wave equation}
\subsection{The improved extended Kudryashov method}
In this section
we will try to use the extended Kudrtashov method to find some general solutions of the Eq. (1).
To obtain the exact solutions of (1+1) dimensional dispersive long-wave equation, we set
\begin{equation}
	\begin{aligned}
	u(x,t)=&A_{0}+ {\sum\limits_{K=1}\limits^{M}}{\sum\limits_{i+j=K}}A_{ij}{\phi}^{i}(\xi){\psi}^{j}(\eta)+{ \sum\limits_{K=1}\limits^{M}}B_{K}{\psi}^{-K}(\eta)+{ \sum\limits_{j=1}\limits^{M}}C_{j}{\psi}^{-j}(\eta),\\
	v(x,t)=&a_{0}+ {\sum\limits_{K=1}\limits^{N}}{\sum\limits_{i+j=K}}a_{ij}{\phi}^{i}(\xi){\psi}^{j}(\eta)+{ \sum\limits_{K=1}\limits^{N}}b_{K}{\psi}^{-K}(\eta)+{ \sum\limits_{j=1}\limits^{N}}c_{j}{\psi}^{-j}(\eta).
	\end{aligned}
\end{equation}
Firstly, we need to determine the values of $M, N$ by homogeneous balance method.
 We can obtain $M=1,N=2$ by caluation. Therefore, we set the following general solution
\begin{equation}
	\begin{aligned}
	&u(x,t)=A_{0}+A_{10}{\phi}(\xi)+A_{01}{\psi}(\eta)+\frac{B_{1}}{{\psi}(\eta)}+\frac{C_{1}}{\phi(\xi)},\\
	&	v(x,t)=a_{0}+a_{10}{\phi}(\xi)+a_{01}{\psi}(\eta)+a_{11}+\frac{b_{1}}{{\psi}(\eta)}+\frac{c_{1}}{\phi(\xi)},
	\end{aligned}
\end{equation}
where $\xi=\kappa_{1}x+\omega_{1}t+r_{1}y,\eta=\kappa_{2}x+\omega_{2}t+r_{2}y$, and  $\kappa_{1}$,$\kappa_{2}$,$\omega_{1},$$\omega_{2}$,$r_{1}$,$r_{2}$,$A_{0}$$,A_{10}$$,A_{01}$$,B_{1}$,$C_{1}$ are underdetermined constants  and the functions
${\phi(\xi)}$ and ${\psi(\eta)}$ satisfy the Bernoulli and the Riccati equations. Substitute (85) into (1), then we will obtain a system of algebraic equations by solving these algebraic equations by Maple. We will get the following traveling wave solution of Eq. (1),
then we can get the exact solution expression of $u$ and $v$ by solving these equations together. Some of the results are shown as follows:
\begin{equation}
	\begin{aligned}
		u_1=&A_0+C_1\frac{R_2+R_1e^{R_1\xi+\xi_0}}{R_1},\\
		v_1=&a_0-\frac{C_1(A_0\kappa_1+\omega_1)}{\kappa_{1}}\frac{R_2+R_1e^{R_1\xi+\xi_0}}{R_1}+c_2(\frac{R_2+R_1e^{R_1\xi+\xi_0}}{R_1})^2,
	\end{aligned}
\end{equation}
where $u=\kappa_1x+\omega_1t,v=\kappa_2x+\omega_2t$.
\begin{equation}
	\begin{aligned}
		u_2=&A_0,\\
		v_2=&a_0+b_1(\frac{R_2c_1\kappa_1}{b_1\kappa_2}\eta+\eta_0)+c_1\frac{R_2+R_1e^{R_1\xi+\xi_0}}{R_1}+c_2(\frac{R_2+R_1e^{R_1\xi+\xi_0}}{R_1})^2,
	\end{aligned}
\end{equation}
where $u=\kappa_1x+\omega_1t,v=\kappa_2x+\omega_2t$.
\begin{equation}
	\begin{aligned}
		u_3=&A_0+C_1\frac{R_2+R_1e^{R_1\xi+\xi_0}}{R_1},\\
		V_3=&a_0+b_1(\frac{2S_2}{\frac{C_1R_2\kappa_1}{\omega_{2}}-\sqrt{\mu}\tanh(\frac{\sqrt{\mu}}{2}\eta+\eta_0)})-\frac{C_1(A_0\kappa_1+\omega_1)}{\kappa_1}\frac{R_2+R_1e^{R_1\xi+\xi_0}}{R_1}+c_2(\frac{R_2+R_1e^{R_1\xi+\xi_0}}{R_1})^2,
	\end{aligned}
\end{equation}
where $u=\kappa_1x+\omega_1t,v=\kappa_2x+\omega_2t$.
\begin{equation}
	\begin{aligned}
		u_4=&A_0+C_1\frac{R_2+R_1e^{R_1\xi+\xi_0}}{R_1},\\
		v_4=&a_0+b_1(\frac{R_2C_1\omega_1}{b_1\kappa_2}\eta+\eta_0)+c_1\frac{R_2+R_1e^{R_1\xi+\xi_0}}{R_1}+c_2(\frac{R_2+R_1e^{R_1\xi+\xi_0}}{R_1})^2,
	\end{aligned}
\end{equation}
where $u=\kappa_1x+\omega_1t,v=\kappa_2x+\omega_2t$.
\begin{equation}
	\begin{aligned}
		u_5=&A_0,\\
		v_5=&a_0-a_{01}\frac{1}{\frac{R_2c_1\kappa_1}{a_{01}\kappa_2}\eta+\eta_0}+c_1\frac{R_2+R_1e^{R_1\xi+\xi_0}}{R_1}+c_2(\frac{R_2+R_1e^{R_1\xi+\xi_0}}{R_1})^2.
	\end{aligned}
\end{equation}
where $u=\kappa_1x+\omega_1t,v=\kappa_2x+\omega_2t$.
\subsection{Hyperbolic tangent method}
In this section we shall seek a solution to Eq. (1) in terms of hyperbolic tangent
functions of the following form
\begin{equation}
	\begin{aligned}
			u(\xi)=&a_0+a_1\tanh\xi,\\
			v(\xi)=&b_0+b_1\tanh\xi+b_2(\tanh\xi)^2
	\end{aligned}
\end{equation}
where $\xi=x-\mu t$ and $a_0,a_1,b_0,b_1,b_2,\mu$ are constants to be determined. Bring (91) into Eq. (1) and set the coefficients of the same power of $\tanh\xi$ equal to zero the following algebraic equations are obtained
\begin{equation}
	\begin{aligned}
		&-\frac{64}{3}a_1-8\mu b_1+8a_1b_0+8b_1a_0-24a_1b_2=0,\\
		&-4\mu b_1-8\mu b_2+4a_1b_0+8a_1b_1+12a_1b_2+4a_0b_1+8b_2a_0+\frac{16}{3}a_1=0,\\
		&-4\mu b_1+8\mu b_2+4a_1b_0-8a_1b_1+12a_1b_2+4a_0b_1-8b_2a_0+\frac{16}{3}a_1=0,\\
		&4a_0a_1-4\mu a_1+4b_1+4a_1^2+8b_2+4a_1^2=0,\\
		&4a_0a_1-4a_1^2+4b_1-4\mu a_1-8b_2=0.
	\end{aligned}
\end{equation}
By solving the above system, we get exact solutions of Eq. (1)
\begin{equation}
	\begin{aligned}
		&u_1=\mu-\frac{2\sqrt{3}\tanh(\mu t-x)}{3},\\
		&v_1=\frac{2}{3}-\frac{2(\tanh(\mu t-x))^2}{3}.
	\end{aligned}
\end{equation}
\begin{figure}[htbp]
	\centering
	\subfigure{}{
		\begin{minipage}[t]{0.24\linewidth}
			\centering
			\includegraphics[width=1.2in]{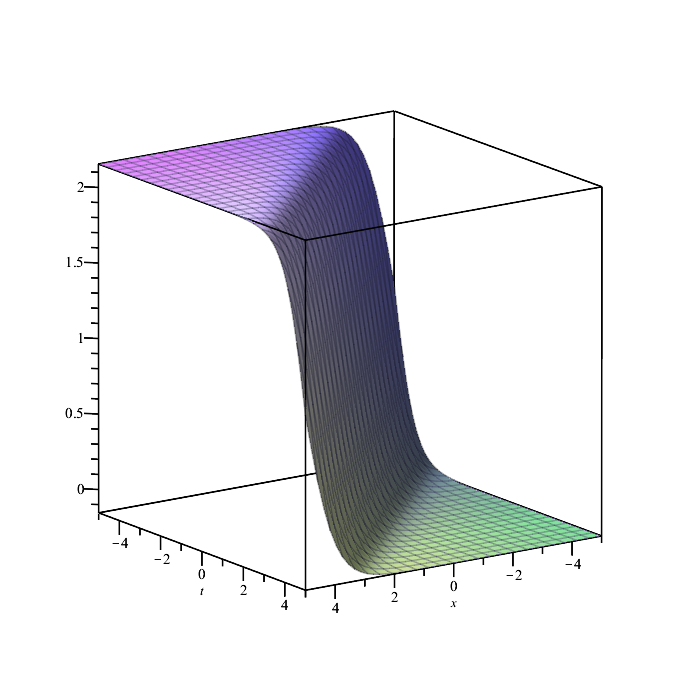}
			%\caption{fig 1}
		\end{minipage}%
	}%
	\subfigure{}{
		\begin{minipage}[t]{0.24\linewidth}
			\centering
			\includegraphics[width=1in]{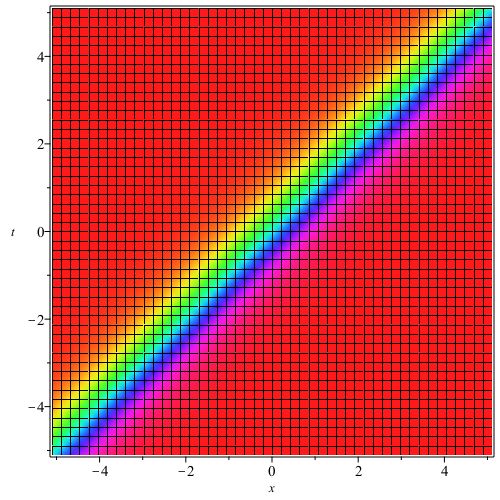}
			%\caption{fig 1}
		\end{minipage}%
	}%
	\subfigure{}{
		\begin{minipage}[t]{0.24\linewidth}
			\centering
			\includegraphics[width=1.2in]{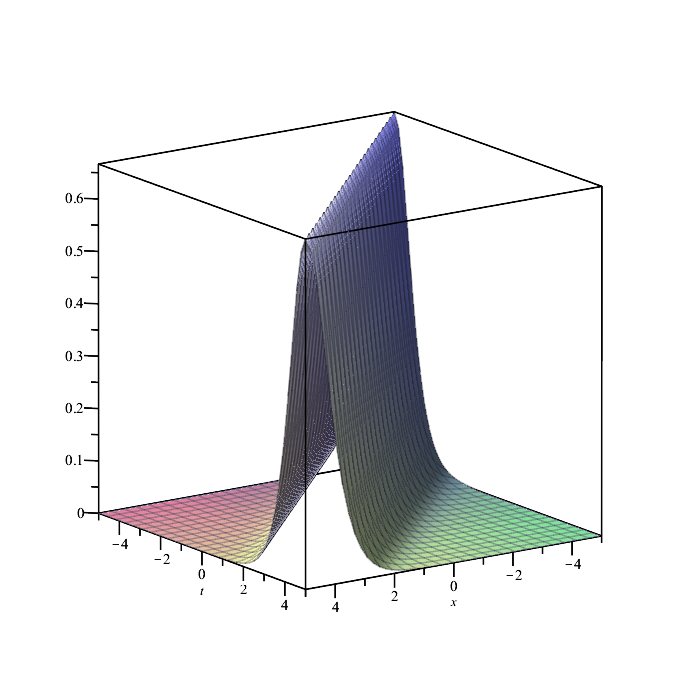}
			%\caption{fig 1}
		\end{minipage}%
	}%
	\subfigure{}{
		\begin{minipage}[t]{0.24\linewidth}
			\centering
			\includegraphics[width=1in]{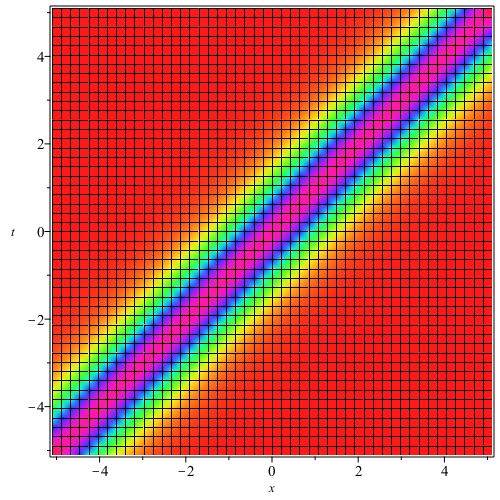}
			%\caption{fig 1}
		\end{minipage}%
	}%
	\centering
	\caption{3D plots and density plots of the $u(x,y,t)$ given by Eq. (138) for parameters $a_1 = 1, t = 1$ and  $a_1 = -1, t = 1$.}
	\label{Fig1}
\end{figure}
\begin{figure}[htbp]
	\centering
	\subfigure{}{
		\begin{minipage}[t]{0.24\linewidth}
			\centering
			\includegraphics[width=1.2in]{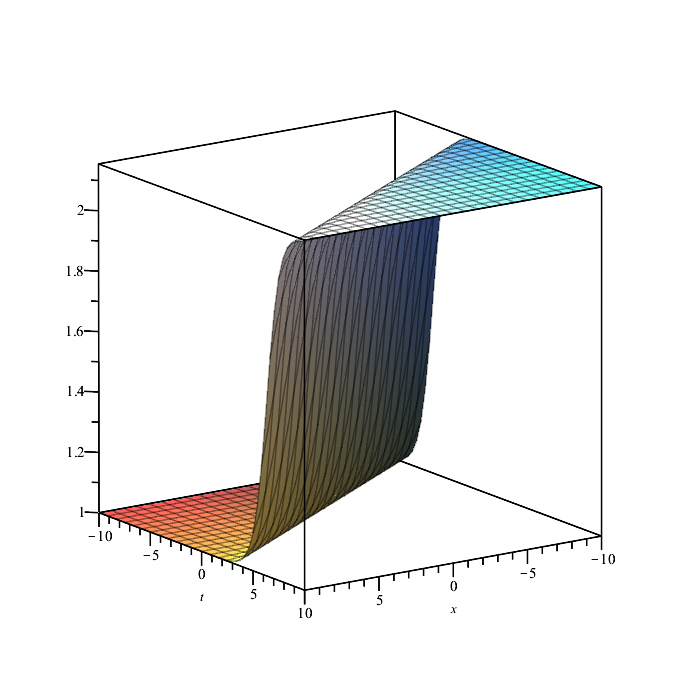}
			%\caption{fig 1}
		\end{minipage}%
	}%
	\subfigure{}{
		\begin{minipage}[t]{0.24\linewidth}
			\centering
			\includegraphics[width=1in]{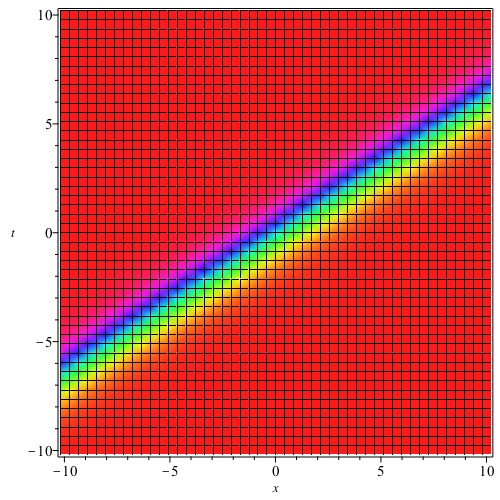}
			%\caption{fig 1}
		\end{minipage}%
	}%
	\subfigure{}{
		\begin{minipage}[t]{0.24\linewidth}
			\centering
			\includegraphics[width=1.2in]{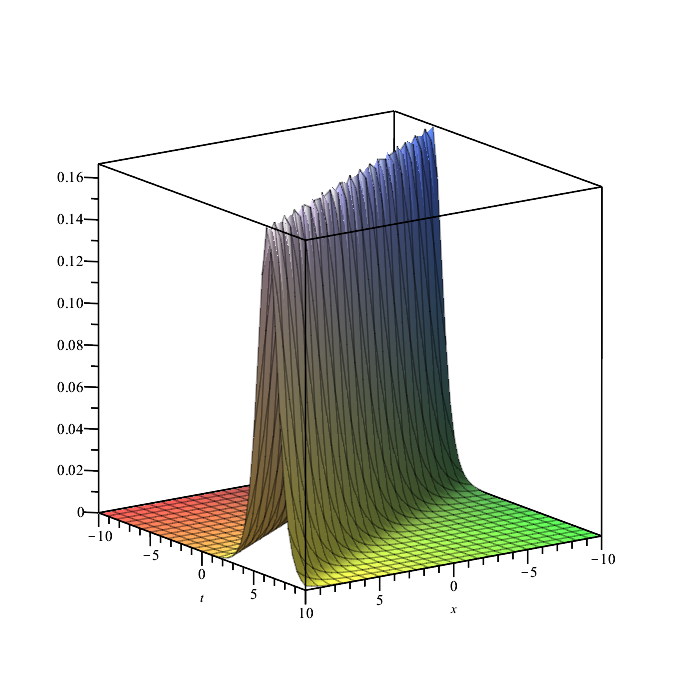}
			%\caption{fig 1}
		\end{minipage}%
	}%
	\subfigure{}{
		\begin{minipage}[t]{0.24\linewidth}
			\centering
			\includegraphics[width=1in]{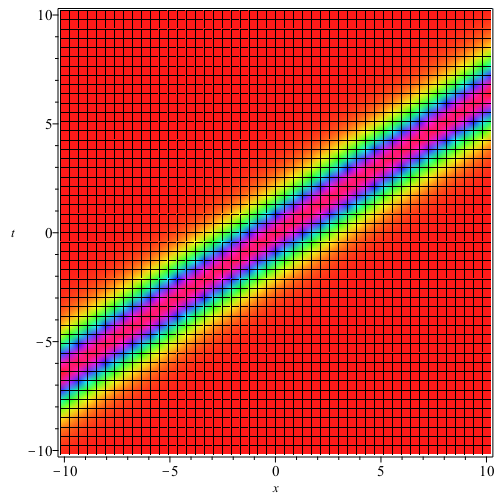}
			%\caption{fig 1}
		\end{minipage}%
	}%
	\centering
	\caption{3D plots and density plots of the $u(x,y,t)$ given by Eq. (138) for parameters $a_1 = 1, t = 1$ and  $a_1 = -1, t = 1$.}
	\label{Fig1}
\end{figure}
\subsection{Rational functions in $\exp(\xi)$}
In this section we shall seek a rational function type of solution to Eq. (1), in terms of $\exp(\xi)$ and we set the following form:
\begin{equation}
	\begin{aligned}
			u(\xi)=&a_0+\frac{a_1}{1+e^\xi},\\
		v(\xi)=&b_0+\frac{b_1}{1+e^\xi}+b_2(\frac{1}{1+e^\xi})^2,
	\end{aligned}
\end{equation}
where $\xi=x-\mu t$ and $a_0,a_1,b_0,b_1,b_2,\omega$ are constants to be determined.
Bring (94) into Eq. (1) and set the coefficients of the same power of $e^\xi$ equal to zero the following algebraic equations are obtained
\begin{equation}
	\begin{aligned}
		&-a_1a_0-a_1^2+\mu a_1-b_1-2b_2=0,\\
		&-a_1a_0+\mu a_1-b_1=0,\\
		&\mu b_1+2\mu b_2-a_1b_0-2a_1b_1-3a_1b_2-b_1a_0-2b_2a_0-\frac{1}{3}a_1=0,\\
		&2\mu b_1+2\mu b_2-2a_1b_0-2a_1b_1-2b_1a_0-2b_2a_0+\frac{4}{3}a_1=0,\\
		&\mu b_1-a_1b_0-b_1a_0-\frac{1}{3}a_1=0. 
	\end{aligned}
\end{equation}
Then the solution may be given as follows
\begin{equation}
\begin{aligned}
	&u_1=a_0+\frac{2\sqrt{3}}{3(1+e^{x-(a_0+\frac{\sqrt{3}}{3})t})},\\
	&v_1=\frac{2}{3(1+e^{x-(a_0+\frac{\sqrt{3}}{3})t})}-\frac{2}{3(1+e^{x-(a_0+\frac{\sqrt{3}}{3})t})^2}.
\end{aligned}
\end{equation}

\section{Conclusion}
In this paper, we first analyze the Lie symmetry of the (1+1) dimensional dispersive long-wave equation, and obtain some Lie point symmetries, invariant solutions and a reduced equation. At the same time, three different methods are used to solve the conservation laws of the ZK equation. The first is to use the direct construction method to solve the multiplier of the target equation.  The adjoint symmetry of the equation is used to substitute the 
symmetry in a partial differential equation that lacks variational symmetry. At this time, the adjoint symmetry satisfies the linear adjoint symmetry determining equations. The
symmetry invariant condition is replaced by the adjoint symmetry invariant condition, and an adjoint symmetry 
formula is presented. The second method is the adjoint equation method. The adjoint equation method proposed by Ibragimov can use the symmetry of
the equation to calculate the conservation law through the explicit formula. It is convenient to calculate
and does not require complex analysis. The third method is to use Noether's theorem. The conservation law is solved by the variational symmetry of the target equation. 
\par In fact, the three methods used in this paper to calculate the conservation law of equations have different advantages. The adjoint equation method proposed by Ibragimov and Noether's theorem can use the symmetry of the equation to calculate the conservation law through the explicit formula. The former is convenient to calculate and does not require complex analysis. It has a wide range of applications, but the results are directly affected by the symmetry of the equation. And the limitation of using the Noether's theorem to solve conservation laws is that sometimes the Euler equation of the target equation cannot be obviously obtained. The advantage of constructing conservation laws directly is that it is not necessary to use the variational symmetry of the equation. For a partial differential equation without variational symmetry, the adjoint symmetry of the equation is used to replace the symmetry. At this time, the adjoint symmetry satisfies the linear adjoint symmetry determining equations. The symmetry invariant condition is replaced by the adjoint symmetry invariant condition, and a formula using adjoint symmetry is given. However, this method is computationally complex and does not apply to any type of equation or equation system. The three methods can be naturally applied to higher dimensional differential equations and differential equation systems.
\par At the same time, we use the symmetry and adjoint symmetry of the obtained target equation to calculate a Lie bracket on the set of adjoint-symmetries given by the range of a symmetry action. Then we use the improved Extended Kudryashov method to solve some exact solutions of (1+1) dimensional dispersive long-wave equation and seek a few solutions in terms of hyperbolic tangent functions and $\exp(\xi)$. In the last part, we analyze the Hamiltonian structure of the target equation, calculate its generalized pre-symplectic operator and solve the linear soliton solution of the target equation.

	\section*{Acknowledgments}
	
	This work is supported by the National Natural Science Foundation of China (Grant No. 12371256 \& No. 11971475).
	
	\section*{Conflict of interest}
	
	The authors declare that they have no known competing financial interests.
	
	\section*{Data availability}
	No data was used for the research described in the article.

	\bibliographystyle{unsrt}
\end{document}